\documentclass[12pt]{article}
\usepackage{epsfig}
\usepackage{enumitem}

\def\beq{\begin{equation}}
\def\eeq#1{\label{#1}\end{equation}}
\def\eeqn{\end{equation}}

\def\beqa{\begin{eqnarray}}
\def\eeqa#1{\label{#1}\end{eqnarray}}
\def\eeqan{\end{eqnarray}}

\let\bar=\overbar

\def\Dslash{\not{\hbox{\kern-4pt $D$}}}
\def\dslash{\not{\hbox{\kern-2pt $\del$}}}

\def\msb{{\bar{\ssstyle M \kern -1pt S}}}

\def\Title#1{\begin{center} {\Large {\bf #1} } \end{center}}

\begin{document}

\Title{Semileptonic Charm Decays: mini review
\footnote{Proceedings of CKM 2012, the 7th International Workshop on the CKM Unitarity Triangle, University of Cincinnati, USA, 28 September - 2 October 2012 }
}

\bigskip\bigskip

%+\addtocontents{toc}{{\it D. Reggiano}}
%+\label{ReggianoStart}

\begin{raggedright}  

{\it Chunlei Liu\index{Liu, C}\\
Department of Physics\\
Carnegie Mellon University\\
Pittsburgh, PA, USA,15213}
\bigskip\bigskip
\end{raggedright}

\section{\boldmath Introduction}

One important task in the field of flavor physics is to over-constrain the CKM matrix.
To best extract CKM matrix elements, we need inputs from both 
experimental data and theoretical calculations. Latest knowledge about the CKM matrix\cite{ckm} shows our understanding
about sin$2\beta$ is the best thanks to clean theory and large statistics from experiments. In contrast, our understanding of both 
mixing and $|V_{ub}|$ have limitations from theoretical predictions.

For example, $|V_{ub}|$ is measured from
$B^0\rightarrow \pi e \nu$ which has the differential decay rate as the following:
\begin{equation}
\frac{d\Gamma}{dq^2} \;=\;  \frac{G^2_F}{24\pi^3}|V_{ub}|^2 p^3_{\pi} |f_+(q^2)|^2.
\end{equation}
In order to extract the value of $|V_{ub}|$ from the branching fraction measurement, we need information on the hadronic form factor $f_+(q^2)$. 
Our best understanding about $f_+(q^2)$ is from lattice QCD calculations. 
To validate improved lattice QCD calculations for the 
form factor in semi-leptonic $B$ decays, we can use semi-leptonic $D$ decays 
as a test and calibration. Since the CKM matrix elements $|V_{cs(d)}|$ can be obtained precisely from
unitarity, we have a reliable method to check the lattice QCD calculations.  

The experimental results of $D$ semi-leptonic decays can be summarized in two main categories: the exclusive decays and the inclusive decays. 
Latest measurements from the exclusive semi-leptonic $D$ decays are discussed here.

\section{\boldmath Measurements of $D \rightarrow K e \nu$ and $D \rightarrow \pi e \nu$}

Semileptonic $D$ decays provide constraints on lattice QCD given CKM matrix elements or vice versa, and they have
been measured in many experiments such as FOCUS, Belle, BaBar and CLEO-c. 
The main goal currently is to improve the precision 
for the Cabibbo-suppressed decay $D \rightarrow \pi e \nu$. 
The BESIII experiment has taken $\sim$2.9 fb$^{-1}$ $\psi(3770)$ data during 
the 2010 and 2011 data runs. Using one-third of the data, preliminary results
 of the $D^0\rightarrow K e \nu$ and $D^0 \rightarrow \pi e \nu$ decays have been obtained.

\subsection{ Event Selection}
The BESIII experiment takes data at the BEPCII symmetric electron-positron collider, with the $\psi(3770)$ produced at threshold.  
Using the double tag technique, several hadronic $D$ decays are fully 
reconstructed at first: $D^0\rightarrow K^-\pi^+$, $D^0\rightarrow K^-\pi^+\pi^0$, $D^0\rightarrow K^-\pi^+\pi^0\pi^0$,
$D^0\rightarrow K^-\pi^+\pi^-\pi^+$.
In case of multiple candidates, the candidate with minimum $\Delta E$ 
is chosen; $\Delta E$ is defined as $\Delta E \;\equiv\;E_{{\rm cand}} - E_{{\rm beam}}$,
where $E_{ {\rm cand}}$ is the energy of the reconstructed tag mode, and $E_{{\rm beam}}$ is the beam energy.
The beam energy constrained mass  $m_{ {\rm BC}}$ is defined as 
$ m_{ {\rm BC}} \;\equiv\; \sqrt{ E^2_{{\rm beam}}-|\vec{p}|^2_{ {\rm cand}}} $,
where $\vec{p}_{ {\rm cand}}$ is the momentum of the reconstructed tag mode.
We require $m_{ {\rm BC}}$ to be between 1.858 GeV and 1.874 GeV before attempting to reconstruct signal candidates.
To search for the signal candidates, the following requirements are applied:
(1) Only two good charged tracks with opposite charges left in the event.
(2) One track is identified as an electron candidate, 
 and the other track is identified as a kaon (pion) candidate for 
 the $K^+ e^- \nu$ ($\pi^+ e^- \nu$) mode. 
(3) The electron has the same charge as the kaon track from the tag side $D$.
(4) The candidate is vetoed if the most energetic unmatched shower has energy greater than 250 MeV, in order to suppress backgrounds with extra $\pi^0$s.
The energy and momentum of the missing neutrino is then inferred by using:
$E_{ {\rm miss}} \;=\; E_{{\rm beam}}- E_{{\rm hadron}}- E_{{\rm electron}}$, and 
$\vec{P}_{ {\rm miss}} \;=\; - \vec{P}_{{\rm tag}} -\vec{P}_{{\rm hadron}} - \vec{P}_{{\rm electron}}$.
The number of signal events are obtained by fitting the $U\equiv E_{ {\rm miss}}- P_{ {\rm miss}}$ distributions.

\subsection{Measurements of Branching Fractions and Form Factors}

Given the signal yields obtained from fitting data and signal efficiencies obtained from signal Monte Carlo, the branching fractions are obtained through the 
following equation:
\begin{equation}
B_{sig} \;=\;  \frac{N^{obs}_{sig}}{\sum_{\alpha} N^{obs,\alpha}_{tag} \epsilon^{\alpha}_{tag,sig}/\epsilon^{\alpha}_{tag} },
\label{eqn:bf}
\end{equation}
where $N^{obs}_{sig}$ is the total number of signal yields with all tag modes combined, $N^{obs,\alpha}_{tag}$ is the observed tag yields for tag mode $\alpha$,
$\epsilon^{\alpha}_{tag}$ is the tag efficiency for mode $\alpha$, and $\epsilon^{\alpha}_{tag,sig}$ is the combined signal and tag efficiency for mode $\alpha$. 
Preliminary results of branching fractions are listed in Table.~\ref{tab:bf}.

\begin{table}[htb]
\begin{center}
\begin{tabular}{l|ccc}  
 mode &  Measurement(\%) &  PDG value (\%) & CLEO-c value (\%) \\ \hline
 $\bar{D}^0\rightarrow K^+ e^- \bar{\nu} $ &   3.542$\pm$0.030$\pm$0.067     &     3.55$\pm$0.04     &    3.50$\pm$0.03$\pm$0.04  \\
 $\bar{D}^0\rightarrow \pi^+ e^- \bar{\nu} $ &   0.288$\pm$0.008$\pm$0.005     &    0.289$\pm$0.008     &  0.288$\pm$0.008$\pm$0.003  \\
 \hline
\end{tabular}
\caption{Branching fraction measurements using $\sim923$ pb$^{-1}$ of $\psi(3770)$ data from BESIII.}
\label{tab:bf}
\end{center}
\end{table}

In order to measure form factor, partial decay rates are measured in different $q^2$ bins, where $q^2$ is the invariant mass squared of the electron-neutrino system.
$ \bar{D}^0\rightarrow K^+ e^- \bar{\nu} $ candidates are divided into nine $q^2$ bins (in GeV$^2$): [0.0,0.2), [0.2,0.4), [0.4,0.6), [0.6,0.8), [0.8,1.0), [1.0,1.2), 
[1.2,1.4), [1.4,1.6), [1.6,$\infty$) , while $\bar{D}^0\rightarrow \pi^+ e^- \bar{\nu} $ are divided into seven $q^2$ bins: [0.0,0.3), [0.3,0.6), [0.6,0.9), [0.9,1.2), [1.2,1.5), [1.5,2.0), 
[2.0,$\infty$). Signal yields in each $q^2$ bin are obtained by fitting $U$ distributions in that $q^2$ range. Using an efficiency matrix vs. $q^2$, obtained
from Monte-Carlo simulation, to correct for smearing, and combining with 
tag yields and tag efficiencies from previous studies, 
the partial decay rates are obtained.

Three different parameterizations of the form factor $f_{+}(q^2)$ are considered,
which are usually referred as the simple pole model~\cite{ff-singlepole},
the modified pole model~\cite{ff-singlepole}, and the series expansion~\cite{ff-series}.

With the above three parameterizations, the data is fitted by minimizing this 
$\chi^2$ function:
% RAB: I changed g(q^2)_i to Delta G_i
% RAB:  more consistent with \Gamma, and also it depends only on 
% RAB:  the bin number,i, and not q^2 directly !
\begin{eqnarray}
\chi^2 &  \,\,=\,\, &  \sum^{n}_{i,j=1} \,\,
       \left( \Delta\Gamma_i - \Delta G_i \right) \,\, C^{-1}_{ij} \,\, 
       \left (\Delta\Gamma_j - \Delta G_j \right),
\end{eqnarray}
where $\Delta\Gamma_i$ is the measured partial decay rate in $q^2$ bin $i$, 
$\Delta G_i$ is the predicted partial decay rate in $q^2$ bin $i$, 
and $C$ is the covariance matrix of the measured partial decay rates.
The fitted results are listed in Table.~\ref{tab:formfit}.

\begin{table}[htb]
\begin{center}
\begin{tabular}{l|c|c|c}
\hline  
Simple Pole &  $f_+(0)|V_{cd(s)}|$  &  $m_{pole}$  &  \\
 $\bar{D}^0\rightarrow K^+ e^- \nu$ &   0.729$\pm$0.005$\pm$0.007     &    1.943$\pm$0.025$\pm$0.003    &     \\
 $\bar{D}^0\rightarrow \pi^+ e^- \nu$ & 0.142$\pm$0.003$\pm$0.001     &    1.876$\pm$0.023$\pm$0.004    &     \\
\hline
Modified Pole &  $f_+(0)|V_{cd(s)}|$  &  $\alpha $  &  \\
 $\bar{D}^0\rightarrow K^+ e^- \nu$ &   0.725$\pm$0.006$\pm$0.007     &    0.265$\pm$0.045$\pm$0.006    &     \\
 $\bar{D}^0\rightarrow \pi^+ e^- \nu$ & 0.140$\pm$0.003$\pm$0.002     &    0.315$\pm$0.071$\pm$0.012    &     \\
\hline
2 par. series &  $f_+(0)|V_{cd(s)}|$  &  $r_1 $  &  \\
 $\bar{D}^0\rightarrow K^+ e^- \nu$ &   0.726$\pm$0.006$\pm$0.007     &    -2.034$\pm$0.196$\pm$0.022    &     \\
 $\bar{D}^0\rightarrow \pi^+ e^- \nu$ & 0.140$\pm$0.004$\pm$0.002     &    -2.117$\pm$0.163$\pm$0.027    &     \\
\hline
3 par. series &  $f_+(0)|V_{cd(s)}|$  &  $r_1 $  &  $r_2$ \\
 $\bar{D}^0\rightarrow K^+ e^- \nu$ &   0.729$\pm$0.008$\pm$0.007     &    -2.179$\pm$0.355$\pm$0.053    & 4.539$\pm$8.927$\pm$1.103     \\
 $\bar{D}^0\rightarrow \pi^+ e^- \nu$ & 0.144$\pm$0.005$\pm$0.002     &    -2.728$\pm$0.482$\pm$0.076    &  4.194$\pm$3.122$\pm$0.448   \\
\hline
\end{tabular}
\caption{
Fitter parameters from form factor measurements for $\bar{D}^0\rightarrow K^+ e^- \nu$ and $\bar{D}^0\rightarrow \pi^+ e^- \nu$.
}
\label{tab:formfit}
\end{center}
\end{table}

\section{Measurement of $D^+\rightarrow K^- \pi^+ e^+ \nu_e$}
\label{sec:vector}

The $D^+\rightarrow K^- \pi^+ e^+ \nu$ decay includes contributions from 
different $K\pi$ resonances, including the dominant $\bar{K}^*(892)^0$ 
piece, as well as a non-resonant amplitude.  
The measurement of the $\bar{K}^*(892)^0$ (lowest mass vector meson) 
contribution can be used to study hadronic transition form factors. Using 347.5 fb$^{-1}$ of data recorded at the $\Upsilon(4S)$ 
the BaBar collaboration has analyzed this decay channel~\cite{paper-barbar}.

\subsection{ Event Selection}
To reconstruct $D^+$ decays, all the charged and neutral particles are boosted to the center-of-mass system and the event thrust axis is determined.  
A plane perpendicular to this axis defines two hemispheres, and the candidate hemisphere consists of a positron, a charged kaon, and a charged 
pion. A vertex fit is performed and events with probability larger than 10$^{-7}$ are kept.  To estimate the neutrino momentum,
the $K^- \pi^+ e^+ \nu_e$ system is constrained to the $D^+$ mass with estimates of the $D^+$ direction and neutrino energy obtained
from all tracks and unmatched showers measured in the event.  
The neutrino energy is evaluated by subtracting from the hemisphere energy 
the energy of reconstructed particles contained in that hemisphere.  
To further reject $B\bar{B}$ background and continuum background 
(arising mainly from charm particles), 
Fisher discriminants are created from relevant variables.  This discriminant retains 40\% of signal events and rejects 94\% of the remaining background.
About 244$\times 10^3$ signal events are selected with a final signal-to-background ratio of 2.3\,.

\subsection{\boldmath Measurement of Form Factors and $\bar{K}^*(892)^0$ Properties}

Form factors $F_{1,2,3}$ can be expanded into partial waves to show explicit dependence on $\theta_K$, 
where $\theta_K$ is the angle between the kaon
three-momentum in the kaon rest frame and the
line of flight of the kaon-pion system in the $D$ rest frame.
 Considering only
$S$, $P$ and $D$ waves, this gives:
$F_1 \;=\;  F_{10} + F_{11}\cos\theta_K +  F_{12} \frac{3\cos^2\theta_K -1}{2},\\  
F_2 \;=\;  \frac{1}{\sqrt{2}} F_{21} +  \sqrt{\frac{3}{2}} F_{22} \cos\theta_K, 
F_3 \;=\;  \frac{1}{\sqrt{2}} F_{31} +  \sqrt{\frac{3}{2}} F_{32} \cos\theta_K 
$,
where $F_{10}$ characterizes the $S$-wave contribution, and $F_{i1}$ and $F_{i2}$ correspond to the $P-$ and $D$-waves, respectively.
It is also possible to relate these form factors with the helicity form factors $H_{0,\pm}$, which can be in turn related 
to the two axial-vector form factors $A_{1,2}(q^2)$ and the vector form factor $V(q^2)$. For the $q^2$ dependence, the simple pole
parameterization is used:
$$
V(q^2)   \;=\;  \frac{V(0)}{1-q^2/m^2_V}  , A_1(q^2) \;=\;  \frac{A_1(0)}{1-q^2/m^2_A}  , A_2(q^2) \;=\;  \frac{A_2(0)}{1-q^2/m^2_A}  
$$
where $m_V$ and $m_A$ are expected to be close to $m_{D^*_s} = 2.1$ GeV and $m_{D_{s1}}=2.5$ GeV, respectively. In the analysis, ratios
of these form factors, evaluated at $q^2=0$, $r_{V}= V(0)/A_1(0)$ and $r_{2}= A_2(0)/A_1(0)$ are measured by studying the variation of 
partial decay rate versus kinematic variables; $m_A$ is allowed to fit while $m_V$ is fixed. For the mass dependence, in case of the 
$K^*(892)$, a Breit-Wigner distribution is used.  
Table.~\ref{tab:barbar1} shows the fit results considering three different models: (1)
a signal made of the $\bar{K}^*(892)^0$ and $S$-wave components; (2)
a signal made of the $\bar{K}^*(892)^0$,  $\bar{K}^*(1410)^0$ and $S$-wave components;
(3)a signal made of the $\bar{K}^*(892)^0$,  $\bar{K}^*(1410)^0$,  $S$- and $D$-wave components.
The fractions of signal components in different models are measured and shown in Table.~\ref{tab:barbar2}. The second model is considered as the
nominal fit to data, with systematics error also listed in the table.

\begin{table}[htb]
\begin{center}
\begin{tabular}{l|ccc}  
Variable & $S + \bar{K}^*(892)^0$   &  $S + \bar{K}^*(892)^0 $  & $S + \bar{K}^*(892)^0 $  \\ 
         &                           &  $\bar{K}^*(1410)^0 $    & $\bar{K}^*(1410)^0 + D $  \\ 
\hline

 $m_{K^*(892)}$ (MeV) &   894.77$\pm$0.08     &   895.4$\pm$0.2$\pm$0.2   & 895.27$\pm$0.21   \\
 $\Gamma^0_{K^*(892)}$( MeV) &   45.78$\pm$0.23     &   46.5$\pm$0.3$\pm$0.2   & 46.38$\pm$0.26   \\
 $r_{BW}$ (GeV$^{-1}$) &   3.71$\pm$0.22     &   2.1$\pm$0.5$\pm$0.5   & 2.31$\pm$0.20   \\
 $m_A$ (GeV)          &   2.65$\pm$0.10     &  2.63$\pm$0.10$\pm$0.13   & 2.58$\pm$0.09    \\
 $r_V$                &   1.458$\pm$0.016     &  1.463$\pm$0.017$\pm$0.031   & 1.471$\pm$0.016    \\
 $r_2$                &   0.804$\pm$0.020     &  0.801$\pm$0.020$\pm$0.020   & 0.786$\pm$0.020    \\

\hline
\end{tabular}
\caption{ Values of form factor and $\bar{K}^*(892)^0$ parameters with different models.}
\label{tab:barbar1}
\end{center}
\end{table}

\begin{table}[htb]
\begin{center}
\begin{tabular}{l|ccc}  
Component & $S + \bar{K}^*(892)^0$   &  $S + \bar{K}^*(892)^0 $  & $S + \bar{K}^*(892)^0 $  \\ 
         &                           &  $\bar{K}^*(1410)^0 $    & $\bar{K}^*(1410)^0 + D $  \\ 
\hline
  $S$-wave              &  5.62$\pm$0.14$\pm$0.13   &  5.79$\pm$0.16$\pm$0.15    &  5.69$\pm$0.16$\pm$0.15   \\
  $P$-wave              &  94.38   &  94.21    &  94.12   \\
 $\bar{K}^*(892)^0$   &  94.38   &  94.11$\pm$0.74$\pm$0.75    &  94.41$\pm$0.15$\pm$0.20  \\
 $\bar{K}^*(1410)^0$  &  0       &  0.33$\pm$0.13$\pm$0.19     &  0.16$\pm$0.08$\pm$0.14  \\
  $D$-wave              &  0       &  0                          &  0.19$\pm$0.09$\pm$0.09  \\
\hline
\end{tabular}
\caption{ Fractions (in percent) of signal components with different models.}
\label{tab:barbar2}
\end{center}
\end{table}

\section{\boldmath Measurement of  $D^0/D^+ \rightarrow \rho e \nu$ and $D^+ \rightarrow \omega e \nu$}

Using 818 pb$^{-1}$ of data taken at the $\psi(3770)$, the CLEO-c collaboration has measured branching fraction and form factor for 
the decays of  $D^0/D^+ \rightarrow \rho e \nu$ and the branching fraction 
for $D^+ \rightarrow \omega e \nu$~\cite{paper-cleoc}. 
The precision of the branching fractions is improved, and the form factor result of the $D^0/D^+ \rightarrow \rho e \nu$ 
is the first measurement on the Cabibbo-suppressed 
pseudo-scalar meson to vector meson transition in semi-leptonic $D$ decay. 

\subsection{Event Selection}
The double-tag technique is used in this analysis by reconstructing a $D$ tag in the following hadronic final states: 
$K^+\pi^-$, $K^+\pi^-\pi^0$, and $K^+\pi^-\pi^-\pi^+$ for neutral tags, and $K^0_s\pi^-$, $K^+\pi^-\pi^-$,  $K^0_s\pi^-\pi^0$, $K^+\pi^-\pi^-\pi^0$, $K^0_s \pi^-\pi^-\pi^+$,  
and $K^-K^+\pi^-$ for charged tags. In case of multiple candidates in the same tag mode, the candidate with minimum $\Delta E$ is chosen. Once a tag is
identified, certain $\Delta E$ and $m_{ {\rm BC}}$ cuts are required. The unused tracks and showers are then searched for a candidate $e^+$ along with a $\rho^-(\pi^-\pi^0)$, $\rho^0(\pi^+\pi^-)$,
or $\omega(\pi^+\pi^-\pi^0)$. The $\rho$ candidate is required to have invariant mass within 150 MeV from the nominal PDG mass. The combined tag and semi-leptonic 
% RAB: added "and unmatched showers" here...
candidates must account for all reconstructed tracks and unmatched showers in the event. To remove multiple candidates in each semi-leptonic mode, one combination is chosen
per tag mode per tag charge, based on the proximity of the invariant masses of the $\rho^0$, $\rho^+⁺$ or $\omega$ candidates to their PDG masses.
 The number of semi-leptonic decays are obtained by fitting the $U\equiv E_{ {\rm miss}}- |\vec{p}_{ {\rm miss}}|$
distributions, where $E_{ {\rm miss}}$ and $\vec{p}_{ {\rm miss}}$ are energy and momentum of the missing neutrino, and they can be inferred from all other measured particles.

\subsection{Measurements of Branching Fraction and Form Factor}
To measure the absolute branching fraction, Monte-Carlo samples are used to determine tag and signal efficiencies. Together with the tag and signal yields,
the same equation as Equation~\ref{eqn:bf} can be used to give the branching fraction. The measured results are listed in Table.~\ref{tab:bfrho}.

\begin{table}[htb]
\begin{center}
\begin{tabular}{l|ccccc}  
Decay mode & $\epsilon$(\%)   &  $N_{sig}$  & $B_{SL}$  & $B_{SL}$(ISGW2)   & $B_{SL}$(FK)  \\
 
\hline
   $D^0 \rightarrow \rho^- e^+ \nu_e$     &  26.03$\pm$0.02   &  304.6$\pm$20.9    &  1.77$\pm$0.12$\pm$0.10  &  1.0  & 2.0   \\
   $D^+ \rightarrow \rho^0 e^+ \nu_e$     &  42.84$\pm$0.03   &  447.4$\pm$24.5    &  2.17$\pm$0.12$^{+0.12}_{-0.22}$  &  1.3  & 2.5   \\
   $D^+ \rightarrow \omega e^+ \nu_e$     &  14.67$\pm$0.03   &  128.5$\pm$12.6    &  1.82$\pm$0.18$\pm$0.07  &  1.3  & 2.5   \\

\hline
\end{tabular}
\caption{ Branching fractions for  $D^0 \rightarrow \rho^- e^+ \nu_e$, $D^+\rightarrow \rho^0 e^+ \nu_e$  and $D^+ \rightarrow \omega e^+ \nu$, from the CLEO-c analysis and two model predictions: ISGW2\cite{rhopaper1} and FK\cite{rhopaper2}. The uncertainties for signal efficiency $\epsilon $ and signal yields
$N_{sig}$ are statistical only. The efficiency includes the $\rho$ and $\omega$ branching fractions from the PDG.
}
\label{tab:bfrho}
\end{center}
\end{table}
 
A form factor analysis is performed for $D^0/D^+ \rightarrow \rho^-/\rho^0 e^+ \nu_e$ decays.
The mechanism of this decay is similar to the $D^+\rightarrow K\pi e \nu$ discussed in Sec.~\ref{sec:vector}, except only $P$-wave is considered in this case.
Three dominant form factors, two axial and one vector, $A_1$, $A_2$, and $V$, 
are used to describe the hadronic current. A simple pole model is assumed with
the pole mass fixed as $M_{D^*(1^-)} = 2.01$ GeV and $M_{D^*(1^+)} = 2.42$ GeV for the vector and axial form factors, respectively. 
A four-dimensional maximum likelihood fit~\cite{rhoLL} is performed and a simultaneous fit
is made to the isospin-conjugate modes $D^0 \rightarrow \rho^0 e^+ \nu_e$ and $D^+ \rightarrow \rho^- e^+ \nu_e$.  The ratios of form factors evaluated at 
$q^2=0$, $r_V = \frac{V(0)}{A_1(0)}$ and $r_2 = \frac{A_2(0)}{A_1(0)}$ are obtained:  $r_V= 1.48 \pm 0.15\pm0.05$ and $r_2 = 0.83\pm0.11\pm0.04$.
Using $|V_{cd}|=0.2252\pm0.0007$, 
$\tau_{D^0} = (410.1\pm 1.5)\times10^{-15}$ s, 
and $\tau_{D^+} = (1040\pm 7)\times 10^{-15}$ s,  
from PDG 2010, form factor ratios and branching fraction results 
are combined to obtain : $A_1(0)=0.56\pm0.01^{+0.02}_{-0.03}$, $A_2(0) = 0.47\pm0.06\pm0.04$, and $V(0)=0.84\pm0.09^{+0.05}_{-0.06}$.

\section{Summary}

Experimental measurements of semi-leptonic $D$ decays have been studied 
very successful during the past few years, with contributions 
from experiments such as FOCUS, Belle, BaBar and CLEO-c. 
During this presentation, recent results on semi-leptonic $D$ decays 
have been discussed, with topic covering several topical aspects, 
i.e., pseudo-scalar to pseudo-scalar modes,
pseudo-scalar to vector modes. 

Since the start of running in 2008, the newest of these experiments, 
BESIII, has taken about 2.9 fb$^{-1}$ of data at $\psi(3770)$.  
With peak luminosity reaching 
more than $6 \times 10^{32}$ (60\% of the designed luminosity),
BESIII is poised to take more data at $\psi(3770)$ and in the higher $D_s$ 
energy region.  
Using part of the data, BESIII has presented preliminary results of
the $D^0\rightarrow K/\pi e \nu$ decays.
Results from the full dataset and other modes are coming in the near future.

\bigskip

The author thanks CKM 2012 committee for the invitation to the conference and
their great organization.

\end{document}